\documentstyle[11pt,paspconf]{article}

\newcommand{\auth}{\sc}
\newcommand{\kms}{{\rm km}\;{\rm s}^{-1}}
\newcommand{\DtoH}{({\rm D}/{\rm H})}

\begin{document}

\title{Galaxy Structure, Dark Matter, and Galaxy Formation}
\author{David H. Weinberg}
\affil{Ohio State University, Department of Astronomy, 174 W. 18th Ave.,
Columbus, Ohio, 43210, U.S.A.}

\begin{abstract}
The structure of galaxies, the nature of dark matter, and the
physics of galaxy formation were the interlocking
themes of {\it DM~1996: Dark and Visible Matter in Galaxies and
Cosmological Implications}.  In this conference summary report,
I review recent observational and theoretical advances
in these areas, then describe highlights of
the meeting and discuss their implications.  I include as an
appendix the lyrics of {\it The Dark Matter Rap: A Cosmological 
History for the MTV Generation}.
\end{abstract}


\section{Introduction}

What is the structure of galaxies?  
What is the dominant form of dark matter in the universe?
How did galaxies form?
As Paolo Salucci explained in his introductory remarks, an
organizing principle of the DM~1996 meeting was to
emphasize the connections between these three issues,
in particular to place an empirical understanding of galaxy
structure at the heart of studies of dark matter and galaxy formation.
While I do not think that the full story of galaxy formation can simply
be read out of observed galaxy properties,
it was clear during this meeting that advances of the past few years
allow much closer links between theoretical accounts of dark matter and galaxy
formation and observations of the Milky Way, nearby galaxies, and
high redshift objects.  

In the next section, I will describe some of the observational and 
theoretical advances that set the scene for DM~1996.
I will then review some of the meeting highlights and discuss their
implications for the structure and formation of galaxies
and the nature of dark matter.
I will close with a discussion of prospects for observational progress
in the next few years.
To complete my introduction, I would like to list some of
the other questions that I had in mind when listening to the conference
talks and composing my summary report.  
These questions are narrower in scope than the three mentioned above,
but we surely will not feel satisfied with our
understanding of galaxies and dark matter until we can answer them all.

\begin{itemize}
\item
Do the observations interpreted as evidence for dark matter actually
reflect a breakdown of general relativity?
\item
Is baryonic dark matter a dynamically important component of galaxies?
\item
Is there non-baryonic dark matter?
\item
How far do dark halos extend?
\item
What are the profiles and shapes of the luminous and dark mass
distributions in galaxies?
\item
What is the origin of the Tully-Fisher relation?  
\item
What environments did the oldest stars form in?
\item
What are typical merger histories for different galaxy types?
\item
What determines a galaxy's luminosity, morphology, color, surface
brightness, and other properties?
\end{itemize}

\section{Recent Advances}

There have been a number of important advances in the last few years,
both observational and theoretical, on the subjects of galaxy structure,
dark matter, and galaxy formation.  These advances set the scene for DM~1996,
and many of the talks presented reviews and updates of these achievements.

\subsection{Observational Advances}

\begin{itemize}
\item
The Discovery of MACHOs -- Paczynski's (1986) proposal to detect massive
objects in the Galactic halo through gravitational microlensing of
background stars seemed, in 1986, brilliant in principle but implausibly
difficult in practice.  The dramatic success of microlensing searches
is a tribute to the guts, cleverness, and perseverance of the experimenters.
This technique has opened up vast opportunities in the studies of dark
matter, Galactic structure, stellar populations, and stellar evolution.
As discussed by {\auth Freeman}
and in \S 3 below, the 2-year results from the
MACHO experiment are quite surprising, and we do not yet know whether
the search has detected the primary constituents of the Milky Way's
dark halo.
\item
The Systematics of Spiral Rotation Curves -- An epochal advance of
the 1970s, beautifully reviewed here by {\auth Van Albada}, was the realization
that galaxy rotation curves are not Keplerian in their outer parts but flat.
This transformation of views moved the dark matter problem to astronomical
center stage.  Over the past few years we have come to appreciate
that rotation curves are not truly flat after all.  Some rise slowly and
continuously, others rise sharply in the middle and decline in the outer
parts, and the shape of a galaxy's rotation curve correlates strongly
with its luminosity and circular velocity.  These developments do not
eliminate the need for dark matter, but they do make the tales 
of ``conspiracy'' between galaxies' luminous and dark components
more complicated.
Many investigators have contributed
to this new understanding of rotation curves
(see, e.g., Salucci \& Frenk 1989; 
Casertano \& van Gorkom 1991; Persic \& Salucci 1991).
The ``Universal Rotation
Curve'' paper of Persic et al. (1996) brought together an
enormous quantity of data in a systematic and powerful way.
The existence of a universal rotation curve has deep implications that
are not yet fully digested, though a number of talks in DM~1996 gave
evidence that people are grappling with them.
It will be crucial for future observational studies to explore the
limitations of the Persic et al.\ (1996) characterization of rotation
curves by examining the degree of scatter at fixed luminosity,
by correlating deviations with other galaxy properties, and by using
more HI data to better tie down rotation curve shapes in the outer parts.
\item
Satellite and Lensing Evidence for Extended Halos --
Binary galaxy studies have long suggested that dark halos extend 
beyond the radius probed by HI measurements, but the studies of
satellite galaxies by Zaritsky and collaborators
(Zaritsky et al.\ 1993; Zaritsky \& White 1994)
made this result much firmer and more quantitative.
{}From the velocity distribution of satellite companions, they
concluded that the rotation curves of $L_*$ spirals remain
approximately flat out to radii of at least 200 kpc.
More recently Brainerd et al. (1996) reached a similar
conclusion by studying the weak lensing of background galaxies
by foreground dark halos, reviving an approach first tried by 
Valdes et al. (1984).  While their initial analysis yielded only a 
$\sim 2\sigma$ detection of extended halos, this technique appears to have 
great potential for application to larger data sets,
and it could perhaps be extended to study the shapes and profiles
of dark halos.  At DM 1996, {\auth Zaritsky} presented the latest
update of the satellite work, using a sample nearly
double the size of Zaritsky et al.'s (1993).
\item
A Partial Consensus on Galaxy Evolution at $z \leq 1$ --
During the 1980s and early 1990s, the evidence of galaxy evolution
seemed to become more confusing and contradictory with each new paper
on number counts or faint galaxy redshifts.  In the last couple of
years, a unified picture has emerged that seems capable of explaining
a wide range of results from ground-based and HST studies:
the more luminous, redder galaxies 
show little evolution between $z=1$ and the present, but the
population of fainter, bluer galaxies with late-type morphologies
and emission-line spectra evolves very rapidly.
These blue galaxies must have been either more luminous or more numerous
at $z=0.5$ than they are today.  There is still no clear consensus
on what these objects evolve into, though low surface brightness dwarfs 
are one popular candidate.  There is also an intriguing recent study
that challenges the account of early-type galaxies.  Kauffmann et al.
(1996) applied a $V/V_{\rm max}$ test, incorporating passive evolution
effects, to the red galaxies in the Canada-France Redshift Survey
(Lilly et al.\ 1995), and they concluded that 2/3 of the luminous,
early-type galaxies were missing from the sample at $z \sim 1$, either
because they were broken into fainter subunits or because active
star formation had altered their colors.
\item
Metals in the Lyman-alpha Forest --
The traditional view of quasar absorbers had long distinguished 
``metal-line'' systems, associated with high HI column densities,
{}from the lower column density ``Ly$\alpha$ forest'' systems, which
were usually thought to be of primordial composition.  Then came
the Keck telescope, the HIRES spectrograph, and studies
revealing heavy element absorption in at least 50\% of ``forest''
lines with $N_{\rm HI} > 3 \times 10^{14}\;{\rm cm}^{-2}$
(Cowie et al. 1995; Womble et al. 1996; for earlier evidence of
metals in the forest see Meyer \& York 1987).  
Even HIRES has difficulty reaching to lower column
densities, but current results are consistent with a metallicity
$\sim 0.01$ solar in all Ly$\alpha$ forest systems.
If the absorbers reside in galaxies, then metal enrichment is 
no great surprise, but if the simulation-based picture that I
described in my contributed talk is correct, then most of the 
forest lines are produced by intergalactic gas (see Cen et al.\ 1994;
Zhang et al. 1995; Hernquist et al.\ 1996).
In this case the detection of metals implies that the primordial
gas was ubiquitously enriched by early star formation before the
formation of most galaxies.  This conclusion is not implausible,
but it is profound, and it will be interesting to explore its consequences
and to try to test it against observations of low metallicity stars
in the Milky Way.

\end{itemize}

\subsection{Theoretical Advances}

\begin{itemize}
\item
The Cluster Baryon Argument --
This simple but powerful argument, originally due, I believe, to 
Simon White, draws a connection between the baryon density parameter
$\Omega_b$ and the mass density parameter $\Omega$.
In standard scenarios of structure formation, the ratio of 
baryonic mass to total mass within the virial radius of a rich
cluster should be close to the global ratio, and observed ratios
therefore set a lower limit (since some of the baryons could be dark)
to $\Omega_b/\Omega$ (White et al.\ 1993).  Combining with
nucleosynthesis estimates of $\Omega_b$, one obtains an upper limit
on $\Omega$.  As I will discuss in \S 3, this argument has begun
to shake theorists' faith in a critical density universe, and it
raises the stakes in studies of the primordial deuterium abundance.
\item
A Universal Density Profile from Hierarchical Clustering ---
Navarro, Frenk, \& White (1996; hereafter NFW) used high-resolution
N-body simulations to show that collisionless collapse in hierarchical
clustering models produces a universal form for the halo density
profile, $\rho \propto r^{-1}$ in the inner parts bending gradually
towards $\rho \propto r^{-3}$ in the outer regions.  The concentration
of the halo, defined, for example, by the ratio of the scale radius 
where $\rho \propto r^{-2}$ to the virial radius within which the
overdensity is $\sim 200$, depends systematically on cosmological
parameters and on the halo's collapse redshift, leading to a tight
correlation between the amplitude and shape of galaxy rotation
curves.  It is obviously tempting to view the ``Universal Profile'' as the
theoretical counterpart to the ``Universal Rotation Curve'' discussed above,
and the one may well prove to be the explanation of the other.  The combination
of a universal shape with a concentration that depends on collapse
redshift suggests a way to unify the ``violent relaxation'' and
``secondary infall'' accounts of gravitational collapse (White 1996),
and the NFW profile offers a natural way to reconcile strong gravitational
lensing with large X-ray core radii in rich clusters.
As a caveat, I should note that there is
not yet complete consensus within the N-body community on the form
of the density profile or the degree of scatter at fixed mass.
{\auth Navarro} showed new results from this work, which I discuss
briefly below, and {\auth Rix} showed the power of incorporating this
kind of physically motivated halo profile into modeling of observations.
\item
Monte Carlo Models of Galaxy Formation and Evolution ---
Kauffmann et al. (1993) and Cole et al.\ (1994),
building on the ideas of Press \& Schechter (1974), White \& Rees (1978),
and White \& Frenk (1991), developed a remarkably rich, semi-analytic
formalism for modeling galaxy formation in hierarchical clustering
models.  This approach allows the kind of population synthesis and
chemical evolution calculations discussed by {\auth Chiosi} and 
{\auth Matteuci}
to be placed within a realistic framework describing the collapse and
merging of dark matter halos and the cooling of gas within those halos.
The most important qualitative lesson to emerge from these studies
is that a galaxy's star formation history and assembly history may be two
quite different things, as much of the star formation can occur in
sub-units that only later merge to form the final galaxy.
Thus, even though the mass scale of typical collapsed objects
grows with time, a hierarchical model predicts blue dwarf irregulars
and red giant ellipticals because the low mass objects that survive
as such today collapsed relatively recently, while the massive objects
formed by mergers in dense regions, where the first collapse of
star-forming systems happened early.  There are free parameters
and approximations in the Monte Carlo models, but they 
can make contact with a wide variety of observations,
and they provide a theoretical framework within which one can quickly assess 
the effect of changing the cosmological scenario or altering the assumptions 
about star formation
and ``microphysics'' (cooling processes, metal enrichment, feedback, etc.). 
{\auth Frenk}'s review described the Monte Carlo methods and presented
some recent results, and the talks by {\auth Firmani}, {\auth Avila-Reese},
and {\auth Menci} showed that others are making use of this general approach,
developing extensions and finding new applications.
\item
Support for the White/Rees Scenario in Cosmological Simulations ---
Collisionless mergers quickly erase substructure, so a purely gravitational
account of galaxy formation cannot explain the existence of rich clusters
containing dozens or hundreds of galaxies.  White \& Rees (1978)
emphasized this problem and proposed the now canonical solution:
dissipation concentrates the baryons within their dark halos, creating
tightly bound units that can survive long after the halos themselves
merge.  Timescale arguments based on spherical collapse suggest that
the required cooling should occur rapidly enough, and numerical 
simulations have recently achieved the dynamic range and physical 
sophistication needed to demonstrate this process in action, showing
the transition from realistic cosmological initial conditions to 
radiatively cooled gas clumps that have masses and overdensities comparable
to the baryonic components of galaxies (Katz et al. 1992;
Evrard et al. 1994; Summers et al. 1995;
Katz et al. 1996).
{\auth Jenkins} et al.'s poster presented some results from state-of-the-art
cosmological hydrodynamics simulations; with the computational firepower
afforded by parallel machines, one can go directly from the primordial
fluctuations of a cosmological theory to the clustered galaxy distribution,
predicting rather than imposing its relation to the underlying distribution
of mass.  Higher resolution simulations of individual galaxy collapses
reveal the complexity of the galaxy assembly process and suggest
clues to the origin of the Hubble sequence
(e.g., Katz \& Gunn 1991; Katz 1992; Navarro \& White 1994;
Vedel et al. 1994; Steinmetz 1995;
Navarro \& Steinmetz 1996; {\auth Gelato} and {\auth Governato} in these
proceedings).  Numerical simulations have played a valuable
role in calibrating the assumptions used in the Monte Carlo models,
and as the simulations improve the interplay between these
approaches should become even more fruitful.  I suspect (and hope)
that Ly$\alpha$ forest simulations like those described in my
contributed talk will also turn out to be an important theoretical
advance (see Hernquist et al. 1996 and Weinberg et al. 1996 for
written accounts of this work).
They are somewhat off the main theme of this conference,
but they do offer further support for the scenario of hierarchical
structure formation, and if the promising initial results survive
tougher scrutiny, then a new class of observations, high-resolution
quasar spectra, will become available as tests of cosmological models.
\end{itemize}

\section{Some Meeting Highlights}

For anyone who has conferenced with Carlos Frenk in years past,
the most astonishing event of DM~1996 was his declaration
(in the discussion following {\auth Girardi}'s talk) that ``There is no reason
to think that $\Omega$ equals one.''  This comment demonstrates the
power of the cluster baryon argument (see \S 2) to shake views once
stoutly defended --- 
there are no obvious minor fiddles to the standard scenario that
circumvent the argument apart from lowering $\Omega$.

I list below a few of the other results that I considered highlights
of the meeting, in chronological order of presentation.
In the sections that follow, I will discuss implications of
these results and others presented at DM~1996 
for our understanding of galaxy structure,
dark matter, galaxy formation, and the Tully-Fisher relation.

\medskip
\noindent
{\bf Navarro:} Halos in $\Omega=1$ CDM models are too concentrated to
fit the observed rotation curves of 
dwarf irregulars and low surface brightness (LSB) spirals.
Low-$\Omega$ models yield better fits.  Decomposing 
observed rotation curves assuming the NFW profile for the dark halo
collapses most of the 
scatter in a plot of concentration vs.\ circular velocity
($r_s/r_{200}$ vs. $v_{200}$, where $r_s$ is the NFW scale radius,
$r_{200}$ the virial radius, and $v_{200}$ the circular velocity
at $r_{200}$.)

\smallskip
Moore (1994) and Flores \& Primack (1995) had argued that the gently
rising rotation curves of dwarf galaxies were inconsistent with the
$r^{-1}$ central cusp of Dubinski \& Carlberg (1991) and NFW,
emphasizing this discrepancy as a challenge to the hypothesis of
CDM halos.  Navarro showed that the same problem appears for giant LSB
spirals, making it more difficult to invoke exotic feedback effects
as a possible solution.
But he also showed that low-$\Omega$ CDM models produce lower concentration
halos (larger $r_s/r_{200}$ at fixed $v_{200}$), in better agreement with
observed rotation curves, and he argued that slowly rising rotation
curves therefore challenge the $\Omega=1$ CDM model rather than the 
hypothesis of cold, weakly interacting dark matter {\it per se}.
Disk/bulge/halo decomposition of the rotation curves in Navarro's
sample yields little scatter in the relation between concentration
and circular velocity for the parent halos, a hint that the predictions
of NFW's numerical simulations may apply to the observed universe.

\medskip
\noindent
{\bf Freeman:} The best fit to the results of the MACHO experiment,
assuming a ``standard'' Galaxy halo, implies a MACHO mass $m=0.5 M_\odot$
and a MACHO halo fraction $f=0.5$.  MACHOs with 
$3\times 10^{-7}M_\odot < m < 0.03 M_\odot$ have $f<0.2$.

\smallskip
The eight microlensing events observed towards the LMC indicate a 
large population of objects whose nature is quite mysterious.
The mass is right for white dwarfs, but deep HST images
reveal no such population (Flynn et al. 1996),
and chemical evolution constraints make any population of stellar
remnants an unlikely candidate for MACHOs 
(Charlot \& Silk 1995; Adams \& Laughlin 1996).
Given the statistical and systematic uncertainties, the detected
MACHOs could represent a small fraction of the dark halo or the
whole thing.  The absence of short-timescale events rules out sub-stellar
objects as the dominant contributors to the dark halo, at least
if the standard halo model is approximately correct.
Two years of MACHO data have raised numerous questions about the Galaxy's
structure and constituents.  Improved statistics, monitoring of ongoing
events, and searches for microlensing in other galaxies 
(as discussed by {\auth Crotts} and {\auth Gondolo}) should guide us towards
answers over the next couple of years.

\medskip
\noindent
{\bf Tytler:} The deuterium abundance in two high-redshift QSO absorbers
implies a baryon density parameter $\Omega_b = 0.024 \pm 0.002 h^{-2}$.

\smallskip
Tytler reviewed the arguments of Tytler et al. (1996) and
Burles \& Tytler (1996) for a primordial deuterium-to-hydrogen ratio
$\DtoH_P = 2.4 \pm 0.3 \times 10^{-5}$, and he argued that higher
values obtained by some other groups are probably a result of
contamination by intervening HI.  Rugers \& Hogan (1996ab), by contrast,
suggest that $\DtoH \sim 2\times 10^{-4}$ is the primordial value
and that other observations obtain lower ratios because of deuterium
destruction.  Combined with big bang nucleosynthesis arguments,
the low $\DtoH_P$ implies a baryon density about double the value
of $\Omega_b=0.0125 h^{-2}$ found by Walker et al. (1991), 
while the high
$\DtoH_P$ implies about half this density, $\Omega_b \approx 0.006 h^{-2}$.
I don't yet find either set of observational interpretations completely
compelling. 
Tytler's value of $\DtoH_P$ is easier to understand theoretically, at 
least if one is willing to expand the quoted observational error bars
on $(^4{\rm He}/{\rm H})_P$. 
With the Rugers \& Hogan (1996ab) abundance, it is
difficult to explain why $\DtoH$ in the local ISM is more than an order of 
magnitude lower,
difficult to reconcile the cluster baryon argument with the density
parameter $\Omega > 0.2$ suggested by large-scale structure studies,
and difficult reproduce the observed mean opacity of the Ly$\alpha$
forest (Hernquist et al.\ 1996).

\medskip
\noindent
{\bf Fontana:} A multi-color search for high redshift galaxies has 
detected 11 galaxies with $3<z<4$ and 5 galaxies with $z>4$, including
one with $z=4.84$.

\smallskip
The Lyman-break method of finding $z>3$ galaxies has truly come into
its own in the last year now that spectroscopic confirmations show
that a large fraction of candidates are genuine high-$z$ objects.
These searches are finally revealing a population of ``normal'' galaxies
at redshifts previously reached only by quasars and radio galaxies.

\medskip
\noindent
{\bf Theuns:} 40\% of stars in the Fornax cluster are intergalactic.

\smallskip
Direct searches for intergalactic background light in galaxy clusters
are notoriously difficult.  Theuns \& Warren (1996) took the
innovative approach of using narrow-band filters to look for
intergalactic planetary nebulae.  They found 10 strong candidates
in Fornax, and after scaling by the ratio of normal stars to planetary
nebulae, they estimate that 40\% of the cluster's stars lie outside
of galaxies.  Freeman reported that an Australian collaboration has
found a similar result in the Virgo cluster.  These are striking
observational results, though it is not too hard to imagine that
tidal stripping and disruption could produce a large population of
intergalactic stars in clusters.  {\auth Gallagher} emphasized that 
galaxy clusters are hostile environments for dwarf ellipticals.
Perhaps Theuns \& Warren are seeing the remnants of those that did not survive.

\medskip
\noindent
{\bf Zaritsky:} The velocity difference $\Delta V$ between a satellite
and its spiral primary is nearly uncorrelated with the line width $W$
of the primary.

\smallskip
This result was reported in the original Zaritsky et al. (1993) paper,
and it is now confirmed in a larger sample.  In the meantime, a
possible theoretical interpretation has emerged, since the systematic
dependence of halo concentration on circular velocity found by NFW
means that a small range of $v_{200}$ maps into a much larger range
of linewidths measured near the optical radius.  While the number
of satellites is probably too small for a detailed quantitative test,
this qualitative explanation of an otherwise puzzling observation is
one of the best pieces of empirical evidence for the NFW profile.
A similar explanation can be constructed from the Persic et al. (1996)
universal rotation curve, if one extrapolates the rotation curve
trends seen near the optical radius to the outer halo.

\medskip
\noindent
{\bf Olling:} The flaring of the HI disk in NGC 4244 implies that its
dark halo is highly flattened, with an axis ratio of about 5:1.

\smallskip
Disk flaring in highly inclined galaxies is one of the few ways of estimating
halo flattening, and Olling's observations and modeling of NGC 4244
define the state of the art (Olling 1996ab).
His conclusion that the halo is highly
flattened depends on the assumption that the gas velocity dispersion
tensor is isotropic.  The observational data do not provide a way
to test this assumption directly, but Olling argued that it is likely
to be correct because the cloud collision time is a small fraction
of the galaxy's age.  Olling is acquiring data for seven additional
galaxies, so we will soon know whether this striking result is found
in a large fraction of spirals.

\medskip
\noindent
{\bf Rix:} Elliptical galaxies have comparable amounts of luminous and
dark matter within the effective radius.

\smallskip
It has been difficult to establish the presence of dark matter in the
inner parts of ellipticals because the orbit population is not known.
Observations that measure the full line-of-sight velocity distribution
out to several effective radii are finally beginning to crack the problem,
demonstrating that models with constant mass-to-light ratio cannot
fit the data.  Assuming an adiabatically contracted, NFW halo, Rix
finds that the luminous and dark mass are about equal within $R_e$
and that the circular velocity stays roughly constant across the 
luminous-to-dark transition.

\medskip
\noindent
{\bf Frenk:} Monte Carlo models of galaxy evolution in the 
$\Omega=1$ CDM model reproduce the observed redshift distribution 
of $B<24$ galaxies, the evolution of the luminosity function 
seen in the Canada-France Redshift Survey, and
the properties of $z>3$ galaxies studied by Steidel et al. (1996).

\smallskip
For many years, the standard CDM model had a reputation for forming
galaxies much too late.  However, while this model does have a 
great deal of activity at $z<1$, it appears that it fits the
observations in this regime and, more remarkably, also
accounts for the population of
star-forming, Lyman-break galaxies at high redshift.
The same kind of modeling should soon lead to predictions for other
cosmological scenarios.

\medskip
\noindent
{\bf Zabludoff:}
75\% of ``E+A'' galaxies are in the field.

\smallskip
A long-standing question has been whether the E+A phenomenon ---
a spectrum showing a recent but not ongoing starburst superposed
on an old stellar population --- is confined to galaxies in
rich clusters, perhaps implicating interaction with the intracluster
medium as the starburst trigger.  Zabludoff's study (Zabludoff et al. 1996)
answers the question definitively, in the negative,
suggesting that the phenomenon is instead triggered by galaxy-galaxy
interactions.  

\section{The Structure of Galaxies}

What is the structure of the main dynamical components in disk galaxies?
A ``traditional'' account, dating from the 1970s but still widely used,
might include an $R^{1/4}$-law bulge, an exponential disk, and a 
spherical, isothermal, dark halo with a constant density core.
Recent observational and theoretical developments, many of them discussed
at DM~1996, suggest a ``modified traditional'' account:
an exponential bulge ({\auth Broeils}; Courteau et al. 1996), 
an exponential disk, and a moderately flattened, possibly triaxial
halo with an NFW profile adiabatically compressed by the luminous matter.
Of course the NFW profile is only the prediction of a specific class
of theoretical models, but it is important to recognize that existing
data are compatible with this and probably many other forms for the 
halo profile, and that theory gives no particular reason to expect
isothermal halos with flat cores (especially after they have
been gravitationally compressed by the luminous matter).
Conventional halo fitting parameters like the core radius and central
density may have more to do with the history of the field than with
the physics of galaxies.

A few speakers discussed more radical views, such as dark matter in
a highly flattened halo or thick disk ({\auth Olling}; {\auth Pfenniger}),
or galaxies governed by non-Newtonian dynamics with no dark matter
at all ({\auth Rodrigo-Blanco}; {\auth Griv}).  {\auth Van Albada} made the
cautionary point that most high-quality, extended rotation curves can
be explained quite adequately by applying MOND (Milgrom 1983)
to the luminous component and detected HI disk, or by applying
Newtonian gravity to the luminous component and a constant 
multiple ($\sim 10$) of the HI disk.  The assumptions are unconventional,
but these models produce good fits to the data with fewer free
parameters than the traditional bulge/disk/halo decomposition.
{\auth Cot\^e} presented an important class of counterexamples
to the scaled HI fits: in dwarf galaxies whose rotation curves
are measured out to many disk scale lengths, the ratio of the HI
surface density to the inferred mass density plummets beyond the
Holmberg radius.  This result does not rule out the possibility of
a dark matter disk, but it weakens the force of the argument by coincidence.
One-dimensional rotation curves have limited power to constrain 
multi-dimensional, multi-component mass distributions, and it is 
therefore important to pursue complementary constraints from modeling
of bars ({\auth De Battista}), warps ({\auth Kuijken}),
faint stellar halos ({\auth Fuchs}), polar rings ({\auth Eskridge}),
or two-dimensional velocity fields ({\auth Schoenmakers}; 
{\auth Mendes de Oliveira}).

The structure of elliptical galaxies is still more difficult to 
pin down because of the absence of dynamically cold tracers.
Nonetheless, X-ray data, gravitational lensing, kinematic
studies of globular clusters, and rotational velocities of 
occasional HI rings all indicate
the presence of extended dark halos, and stellar dynamical evidence
now indicates comparable amounts of dark and luminous mass within
the effective radius 
({\auth Danziger}; {\auth Rix}; {\auth Saglia}; {\auth Vine}).
Modeling observations with adiabatically squeezed, NFW halos, {\auth Rix}
infers circular velocities $v_{200} \geq 400\;\kms$ for two
$L_*$-ish galaxies, suggesting that the dark halos of ellipticals
are more massive than those of spirals of similar luminosity.
{\auth Prugniel} argued that the small tilt of the fundamental plane
is explained by a combination of changes in stellar population,
rotational support, and profile shape with luminosity, concluding
that the ratio of total mass to visible mass in the inner parts
of ellipticals is independent of luminosity.  The strong evidence
for ``normal'' dark halos in ellipticals argues indirectly against
dark disks in spirals, since the dark matter would need to
be dissipative in one class of galaxies but not in the other.

\section{The Nature of Dark Matter}

What is the dark matter?  Traditional ideas include faint
main-sequence stars, sub-stellar objects, and cold, weakly
interacting particles.  The first of these possibilities has
been ruled out by the paucity of faint stars in deep HST images
(Bahcall et al. 1994; Flynn et al. 1996).
The second now appears to be ruled out by the absence of short-timescale
events in the MACHO data, at least for a standard halo model ({\auth Freeman}).
The third idea, beloved of particle physicists and of cosmological
theorists like myself, is harder to kill.

A number of alternative ideas were discussed at DM~1996.
The MACHO results have revived the popularity of stellar
remnants as a dark matter candidate, but it remains difficult
to see how one could process 90\% of the baryons in the universe
into compact objects without violating 
observational constraints on the chemical abundances of halo stars
and the luminosities of galaxies at high redshift.
{\auth Freeman} suggested that the
MACHO experiment might instead be detecting a population of primordial
black holes.  It would be rather mysterious if objects formed a
fraction of a second after the big bang happened to lie in the
mass range of typical stars, but sometimes a coincidence is just
a coincidence.  {\auth Pfenniger} made an impressive case for fractally
clumped molecular gas as dark matter, and {\auth Field \& Corbelli}'s
poster described some closely related possibilities.
These ideas have the virtue of making some interesting, observationally
testable predictions.  However, while one can easily imagine clumpy
gas hiding itself in spiral disks, it seems less natural as an 
explanation for dark matter in elliptical galaxies or in galaxy clusters.
There is also the long-standing argument that baryonic dark matter
alone cannot explain dynamical estimates of $\Omega$ without doing
violence to big bang nucleosynthesis.  This argument holds even with
{\auth Tytler}'s lower estimate of the primordial deuterium abundance,
though the implied gap between $\Omega_b$ and $\Omega$ is no longer
quite as convincing.

Warm elementary particles were discussed as a possible explanation for
slowly rising rotation curves, though {\auth Navarro} argued that one 
could also match these observations with cold dark matter in a low density
universe.  {\auth Soleng} discussed three rather more exotic possibilities
{}from theoretical particle physics: a cosmological constant, an
oscillating gravitational constant, or a string fluid.
These forms of stress-energy have negative pressure, and by
accelerating the cosmic expansion at late times they would help
reconcile the estimated ages of globular clusters with estimated
values of the Hubble constant.  
The first two would not
cluster gravitationally, so they could not be the dark matter in
galaxies and galaxy clusters, but they could still reconcile a low dynamical
$\Omega$ with the flat universe preferred by inflation models. 
I have previously been rather 
skeptical of the cosmological constant as a solution to the age 
problem --- it seems so cheap --- but there are many possibilities
lurking in string theory and extensions of the standard particle model 
for negative-pressure fields that would act like a time-variable $\Lambda$
(see, e.g., the discussion in Wilczek 1997).  The more I hear
about them, the more my resistance weakens.

There remains the heretical alternative to any form of dark matter,
changing the theory of gravity.  Speaking for myself, I believe
firmly in dark matter six days a week, and on the seventh day, I waver.
We cannot happily dismiss the alternative-gravity hypothesis until
we have detected the dark matter by non-gravitational means
(or by microlensing, since its lengthscale and acceleration regime
is completely different from that in which the dark matter problem
manifests itself).  However, while the direct, high-precision tests
of general relativity are confined to scales much smaller than 
galaxies, the conventional GR $+$ dark matter scenario receives enormous
indirect support on larger scales from its successful account of
gravitational lensing, from the multiple empirical 
successes of the big bang theory,
and from the more limited but still significant success of 
gravitational instability theory in explaining the transition from
microwave background fluctuations to galaxies and large-scale structure.
It seems unlikely that a relativistic generalization of non-Newtonian
gravity would preserve these achievements, though conceivably the
departures from GR could be linked to cosmology in a way that suppresses
their magnitude at early times.
(It is a notable coincidence that the characteristic acceleration in
Milgrom's MOND formulation is approximately equal to the speed of light 
multiplied by $H_0$.)  I think that non-Newtonian gravity received
about the right amount of attention at DM~1996: discussed in a few
talks and posters ({\auth Rodrigo-Blanco}; {\auth Van Albada}; {\auth Griv})
and mentioned in comments and questions, but not a dominating theme.

\section{Galaxy Formation}

How did galaxies form?  Many talks presented numerical or semi-analytic
illustrations of ``the CDM view'' of galaxy formation.
The growth of structure in the dark matter distribution is hierarchical
and complex.  Gas cools and forms stars in sub-galactic and galactic
mass halos.  Halos merge rapidly in groups and clusters, but the denser,
gaseous/stellar components merge more slowly.  Roughly speaking, this
is the White \& Rees (1978) scenario, though there is probably greater
appreciation today of the variance in halo merger histories and of 
the influential role of tidal fields and filaments in organizing and
guiding gravitational agglomeration.  The broad features of this
picture are common to the many post-COBE variations on ``standard'' CDM ---
tilted spectrum, low $\Omega$, mixed dark matter, and so forth ---
and some of these features would probably survive in other cosmological models.

The main point of contention in recent theoretical studies has been the
importance of supernova feedback.  The majority view is that feedback
plays a critical role in suppressing the formation of faint galaxies,
and perhaps in controlling the rates of gas cooling and star formation
and in suppressing the transfer of angular momentum between the baryons
and the dark matter.  Dissenters (e.g., Katz, Hernquist, and I)
have assigned feedback a more limited role, arguing that energy
deposited by supernovae cannot travel far in the dense gaseous environment
of a forming galaxy.  I don't know which of these points of view will
ultimately prove closer to the truth, but just for fun I will state
an unequivocal prediction: when the detailed, compelling theory of
galaxy formation that we are groping towards today emerges in its full glory,
supernova feedback will not play a major role in determining the global
properties of galaxies, except for the lowest mass systems
($v_c \leq 50\;\kms$ in round numbers).

The CDM picture of galaxy formation has some empirical support
{}from observations of high-redshift galaxies (e.g., {\auth Fillipi} et al.;
{\auth Frenk}) and from the aforementioned successes of the NFW profile
in explaining rotation curve shapes and satellite dynamics.
However, one could hardly claim that current observations dictate
such a picture.  The strength of the CDM scenario is that it provides
a plausible account of galaxy formation that is integrated with 
evidence from microwave background anisotropies, large-scale structure,
and the Ly$\alpha$ forest.  It is worth noting that if all dark
matter {\it is} baryonic then we have no good {\it a priori} theory of
galaxy and structure formation.  An attraction of primordial black 
holes as dark matter
is that they could explain the MACHO results while preserving
the successes of the CDM scenario.  Because they would form in the
very early universe, they would allow $\Omega$ to exceed $\Omega_b$
without requiring any new elementary particles.

\section{The Tully-Fisher Relation}

The Tully-Fisher relation lies at the nexus of my three organizing
themes, and it was a recurring motif in many of the talks, posters,
and informal discussions at DM~1996.  Typical estimates of the
Tully-Fisher scatter in large samples are $\sigma \sim 0.35$ mag
(Strauss \& Willick 1995).  For his sample of spirals in Ursa Major,
{\auth Verheijen} finds a scatter of 0.26 mag in the K-band
Tully-Fisher relation, and after accounting for observational errors he
estimates an {\it intrinsic} scatter of just 0.12 mag.
The results of Bernstein et al. (1994) for a sample of 25 spirals in
the vicinity of Coma are even more remarkable: they obtain an
{\it observed} scatter of 0.12 mag, leaving almost no room for
intrinsic scatter at all.

A naive physical interpretation of the Tully-Fisher relation is that
it connects a galaxy's stellar mass (indicated by luminosity) to
the depth of its dark matter potential well (indicated by the 
asymptotic circular speed).  The sign of the correlation makes
perfect sense --- more massive halos form more stars and make
deeper potential wells --- and a primordial power spectrum of index
$n \approx -2$ leads to a relation of about the right form if one
assumes that luminosity is proportional to dark halo mass
(Faber 1982; Gunn 1982).  However, this simple argument from collapse
dynamics fails to explain the small scatter of Tully-Fisher because
perturbations of the same mass that collapse at different times
have different post-collapse densities, and hence different
circular speeds (Cole \& Kaiser 1989; Eisenstein \& Loeb 1996).
Disk asymmetries, warps, and non-circular motions all add to the
scatter, making the problem even worse.  Furthermore,
stellar populations influence the relation between luminosity
and stellar mass, the luminous matter influences the rotation curve,
and rotation curves are not always flat, so even the
low-level interpretation of the relation's physical significance
may be incorrect.

The collapse dynamics derivation explains Tully-Fisher 
in terms of cosmological initial conditions, collisionless dynamics,
and an assumption that the luminosity is proportional to the mass
of baryons in the pre-galactic perturbation.
White \& Frenk (1991) present a derivation based instead on supernova
feedback, with the circular velocity of the dark halo controlling the
rate at which gas is able to cool and form stars.
Alternatively, the luminous component might determine the observed
linewidth by modulating the inner structure of the dark halo.
The last two of these explanations are nearly opposite:
in one the dark matter instructs the baryons how to behave
(circular velocity dictates luminosity), and in the other the
baryons instruct the dark matter how to behave (luminous mass
dictates linewidth).  The fact that three qualitatively different
arguments seem at least somewhat plausible, though none entirely
satisfying, illustrates our state of confusion.  I believe that
when we understand (and know that we understand) the true origin
of the Tully-Fisher relation, we will be much closer to understanding
the essential physical processes that govern disk galaxy formation.

Persic and Salucci's view, as I understand it, is that the Tully-Fisher
relation is not in itself fundamental but is instead the consequence
of more basic correlations, much as the Faber-Jackson relation is now
seen as a projection of the fundamental plane for ellipticals.
{\auth Pharasyn} et al. explicitly set out to find a fundamental
plane for spiral galaxy halos.  I think this point of view is quite
tenable and may well lead to useful insights.  However, if we accept
the common wisdom (and perhaps we should not), then the situation for
ellipticals and spirals is quite different, for the central velocity
dispersion of an elliptical depends mainly on the gravitational potential
of the luminous matter, while the linewidth of a spiral depends 
mainly on the gravitational potential of the dark halo.  In the 
former case we can use the virial theorem to connect the luminosity
directly to the dynamical measure, but in the latter case we cannot.
Furthermore, if the scatter in Tully-Fisher is really as small as
the observations suggest, then describing Tully-Fisher as a consequence
of other correlations merely shifts the problem to explaining why
these other patterns are so regular.  The scatter in a ``halo fundamental
plane'' must be smaller than the Tully-Fisher scatter or it cannot
``explain'' Tully-Fisher in the first place, since projection can 
increase scatter but cannot decrease it.  Even for ellipticals, the 
existence of a fundamental plane with small scatter does not in itself
explain the Faber-Jackson relation, for if the plane were uniformly populated
by galaxies, then any off-axis projection would produce a scatterplot
instead of a correlation.

Tully-Fisher surveys of the galaxy peculiar velocity field usually start with
a sample that is as homogeneous as possible in order to obtain the
smallest possible scatter, and hence the best
possible distance estimates.  For understanding galaxy formation,
however, the observational desiderata in a Tully-Fisher
sample are rather different: one wants a broad spectrum of galaxy
types so that one can measure the full scatter at fixed linewidth
and correlate residuals against
galaxy morphology, surface brightness, color, environment, and so forth.
Many observers are beginning to take this approach to Tully-Fisher
studies and are working hard to procure and analyze the necessary
data.  {\auth Van Albada} showed results from an ambitious survey,
now close to completion, that will measure extended HI rotation curves
and velocity fields for 200 galaxies.  {\auth De Blok} discussed the
Tully-Fisher relation for a sample of LSB spirals
and concluded that they lie on the same relation as high surface brightness
spirals, implying that they have systematically higher mass-to-light ratios
(see Sprayberry et al. 1995; Zwaan et al. 1995; de Blok \& McGaugh 1996).
{\auth Matthews} et al. analyzed a different sample of LSB galaxies
and found that they lie an average of 1.3 mag {\it below} the
normal Tully-Fisher relation.  They suggested that the difference
between their result and those of Sprayberry et al. (1995) and
Zwaan et al. (1995) could be accounted for by the difference
between volume-limited and magnitude-limited galaxy samples,
provided that the {\it scatter} in the Tully-Fisher relation is
larger for LSB galaxies.  {\auth Verheijen} used his beautiful
Ursa Major data set to compare Tully-Fisher relations in B, I, and K
for three different measures of linewidth ($W_{\rm HI}$, $2V_{\rm max}$,
and $2V_{\rm flat}$).  While the combination of K magnitudes and
$V_{\rm flat}$ linewidths yielded the smallest scatter, other
combinations were not radically different.
{\auth Hendry} found that he could obtain a 
modest (but only modest) reduction in Tully-Fisher scatter for a
different data set by using the velocity at 0.65 optical radii
in place of the ``asymptotic'' velocity, or by using a template fit to
the Persic et al. (1996) universal rotation curve.
{\auth Rhee} described a principal component analysis of galaxy
rotation curves that may ultimately lead to new formulations of the
Tully-Fisher relation and new dynamical quantities to correlate
residuals against.  {\auth Bershady} presented Tully-Fisher results
for a sample of galaxies at moderate redshift.
I can't say that this wealth of new observations has yet
led to any grand synthesis (for me, anyway), but it is clear that
the data environment for examining the Tully-Fisher relation
and its implications for galaxy formation is getting much richer.

\section{Observational Prospects}

The key observational results reported at DM~1996 come from an 
impressive variety of instruments and techniques.
The Keck telescope is making dramatic progress on problems
that were previously inaccessible, such as the deuterium
searches described by {\auth Tytler}.
The Hobby-Eberly telescope will join Keck later this year ({\auth Bershady}),
and other large aperture telescopes will follow in the near future.
Big telescopes with high-quality spectrographs are revolutionizing
the study of galaxy evolution, especially in combination with multi-color
selection of high-redshift galaxies ({\auth Fontana}) and with
HST imaging, which can reveal
the morphology of galaxies at intermediate redshifts ({\auth Filippi} et al.)
and measure number counts and angular sizes to extremely faint magnitudes.

In the areas of galaxy structure and dark matter, the most important
results are coming from innovative, ambitious applications of
smaller optical telescopes and aperture synthesis radio arrays.
These include ``traditional'' microlensing searches ({\auth Freeman})
and studies of ``pixel'' microlensing ({\auth Crotts}; {\auth Gondolo}),
quasar microlensing ({\auth Hawkins}), and weak lensing ({\auth Broadhurst}),
which take advantage of the wide fields and monitoring capabilities
afforded by CCD cameras on moderate-sized telescopes.
This list has a rather obvious moral: gravitational lensing 
(including old-fashioned macrolensing, as discussed by {\auth Danziger} and 
{\auth Rix}) is an excellent tool for studying dark matter because
it responds directly to mass.
We will also learn a great deal from extensive, systematic surveys that
combine optical/IR imaging with optical spectroscopy and/or
HI mapping ({\auth Broeils}; {\auth Verheijen}; {\auth van Albada};
{\auth De Blok}).

We have many other observational developments to look forward to in
the next few years.  I have already mentioned the hope for a consensus
on primordial deuterium, and from it a measurement of the cosmic
baryon density.  There is also the possibility that accelerators
or detection experiments will turn up a compelling 
candidate for non-baryonic dark matter.
The study of galaxy structure will be advanced 
enormously by the photometry and spectroscopy of the Sloan Digital
Sky Survey (see Gunn \& Weinberg 1995), which will provide for millions
of galaxies the kind of data that are available for hundreds today.
Better constraints on cosmological parameters and theories of 
structure formation will emerge from giant redshift surveys
(the Sloan survey and the Anglo-Australian, 2-Degree Field survey),
{}from the Ly$\alpha$ forest, and from ground- and balloon-based
observations of microwave background anisotropies.  In the slightly
longer term, the planned microwave background satellites --- if they
produce results that accord with current theoretical
expectations --- will yield precise measurements of cosmological
parameters and initial conditions, and we will be left to work out
how these lead to observed galaxies.

At the end of our meeting, Persic and Salucci suggested that we mark
our calendars for DM 2000.  If the developments of the next four years
are as striking as those of the last four, we can expect either to have
a much firmer understanding of galaxy structure, dark matter, and
galaxy formation, or to be rather deeply puzzled.

\acknowledgments

I am grateful to Massimo Persic and Paolo Salucci for organizing
a superb conference and for inviting me to give this summary report.
I am grateful to all of the participants for providing so much
interesting material to summarize!  I thank Hans-Walter Rix for
many stimulating discussions during the conference, and, above all,
for guiding me on a fabulous pre-conference trip through the Dolomites, 
thus making it possible for me to concentrate on the talks.
I also thank Andy Gould for many enlightening conversations about
MACHOs, MOND, and dark matter.

\section{Appendix: The Dark Matter Rap}

I composed ``The Dark Matter Rap: A Cosmological History for the MTV
Generation'' in the fall of 1992, and I first performed it at the
Institute for Advanced Study's Tuesday Lunch on December 8 of that year.
I have avoided circulating the lyrics in the past mainly because I think the 
piece (I hesitate to call it a ``song'') is much more fun to hear
or to see performed than it is to read.  However, by now I've performed
it {\it ad} sufficient {\it nauseam} that a good fraction of its potential
international audience has heard it at least once.  Since
I did close my conference summary with it and could probably never
find a more appropriate opportunity to publish the text, I have decided to
include it here.  I hope that someday soon I will get
around to installing an audio version on my WWW page
(http://www-astronomy.mps.ohio-state.edu/$\sim$dhw/), so I encourage readers
who haven't heard this folly, live or taped, to try there before
reading any further.  I have resisted the
temptation to include references, since I chose names with as
much attention to rhyme and meter as to historical accuracy.

\medskip

My name is Fritz Zwicky,

I can be kind of prickly,

This song had better start

by giving me priority.

Whatever anybody says,

I said in 1933.

Observe the Coma cluster,

the redshifts of the galaxies

imply some big velocities.

They're moving so fast,

there must be missing mass!

Dark matter.

\medskip
Dark matter: Do we need it?  What is it?  Where is it?  How much?

Do we need it?  Do we need it?  Do we need it?  Do we need it?

\medskip
For nearly forty years,

the dark matter problems sits.

Nobody gets worried 'cause, ``It's only crazy Fritz.''

The next step's not 'til the early 1970s,

Ostriker and Peebles,

dynamics of the galaxies,

cold disk instabilities.

They say: ``If the mass, were sitting in the stars,

all those pretty spirals, ought to be bars!

Self-gravitating disks?  Uh-uh, oh no.

What those spirals need is a massive halo.

And hey, look over here, check out these observations,

Vera Rubin's optical curves of rotation.

They can provide our needed confirmation:

those curves aren't falling, they're FLAT!

Dark matter's where it's AT!

\medskip
Dark matter: Do we need it?  What is it?  Where is it?  How much?

What is it? What is it? What is it? What is it?
\medskip

And so the call goes out for the dark matter candidates:

black holes, snowballs, gas clouds, low mass stars, or planets.

But we quickly hit a snag because galaxy formation

requires too much structure in the background radiation

if there's only baryons and adiabatic fluctuations.

The Russians have an answer: ``We can solve the impasse.

Lyubimov has shown that the neutrino has mass.''

Zel'dovich cries, ``Pancakes!  The dark matter's HOT.''

Carlos Frenk, Simon White, Marc Davis say, ``NOT!

Quasars are old, and the pancakes must be young.

Forming from the top down it can't be done.''

So neutrinos hit the skids, and the picture's looking black.

But California laid-back, Blumenthal \& Primack

say, ``Don't have a heart attack.

There's lots of other particles.

Just read the physics articles.

Take this pretty theory that's called supersymmetry.

What better for dark matter than the L-S-P?

The mass comes in at a  $\sim$keV,

and that's not hot, that's warm.''

Jim Peebles says, ``Warm?  Don't be half-hearted.

Let's continue the trend that we have started.

I'll stake out a position that's bold:

dark matter's not hot, not warm, but COLD.''

Well cold dark matter causes overnight sensations:

hand-waving calculations,

computer simulations,

detailed computations of the background fluctuations.

Results are good, and the prospects look bright.

Here's a theory that works!  Well, maybe not quite.

\medskip
Dark matter: Do we need it?  What is it?  Where is it?  How much?

Where is it?  How much?  Where is it?  How much?
\medskip

We have another puzzle that goes back to Robert Dicke.

Finding a solution has proven kind of tricky.

The CMB's so smooth, it's as if there'd been a compact

between parts of the universe that aren't in causal contact.

Alan Guth says, ``Inflation,

will be our salvation,

give smoothness of the universe a causal explanation,

and even make the galaxies from quantum fluctuations!

There is one prediction, from which it's hard to run.

If inflation is correct, then Omega should be one.''

Observers say, ``Stop, no, sorry, won't do.

Look at these clusters, Omega's point 2.''

The theorists respond, ``We have an explanation.

The secret lies in biased galaxy formation.

We're not short of critical mass density.

Just some regions are missing luminosity.''

Observers roll their eyes, and they start to get annoyed,

But the theorists reply, ``There's dark matter in the voids.''

\medskip
Dark matter: Do we need it?  What is it?  Where is it?  How much?

Do we need it?  Do we need it?  Do we need it?  Do we need it?
\medskip

Along comes Moti Milgrom,

who's here to tell us all:

``This dark matter claptrap 

has got you on the wrong track.

You're all too mired in conventionality,

wedded to your standard theory of gravity,

seduced by the elegance of General Relativity.

Just change your force law, that's the key.

Give me one free parameter, and I'll explain it all.''

``Not so,'' claim Lake, and Spergel, et al.,

``On dwarf galaxies, your theory does fall.''

The argument degenerates; it's soon a barroom brawl.

\medskip
Dark matter: Do we need it?  What is it?  Where is it?  How much?

What is it?  What is it?  What is it?  What is it?
\medskip

New observations hit the theory like an ice cold shower.

They show that cold dark matter has too little large scale power.

Says Peebles: ``Cold dark matter?  My feeblest innovation.

An overly aesthetic, theoretical aberration.

Our theories must have firmer empirical foundation.

Shed all this extra baggage, including the carry-ons.

Use particles we know, i.e., the baryons.

Others aren't convinced, and a few propose a mixture

of matter hot and cold, perhaps with strings or texture.

And nowadays some physicists are beginning to wonder

if it's time to ressurrect Einstein's ``greatest blunder.''

Why seek exotic particles instead of just assume

that the dark matter's all around us -- it's what we call the vacuum?

Who's right?  It's hard to know, 'til observation or experiment

gives overwhelming evidence that relieves our predicament.

The search is getting popular as many realize

that the detector of dark matter may well win the Nobel Prize.

So now you've heard my lecture, and it's time to end the session

with the standard closing line: Thank you, any questions?

\bigskip
\ \phantom{with the standard closing line: Thank you,} - DW, 12/92 

\end{document}